\documentstyle[12pt]{article}
\input{psfig.tex}
\begin{document}

\begin{tabbing}
\hskip 11.5 cm \= {Imperial/TP/96-97/12}
\\
\> November 1996 \\
\> astro-ph/9612017 \\
\end{tabbing}
\vskip 1cm
\begin{center}
{\Large \bf How to falsify scenarios with primordial fluctuations from
inflation}\footnote{To be published in Critical Dialogues in
Cosmology (proceedings of the Princeton 250th Anniversary conference,
June 1996), ed. N. Turok (World Scientific) }
\end{center}
\begin{center}
\vskip 1.2cm
{\large\bf Andreas Albrecht}\\
Blackett Laboratory, Imperial College\\
Prince Consort Road, London SW7 2BZ  U.K.\\
albrecht\@ic.ac.uk
\end{center}
\vskip 1cm

\author{Andreas Albrecht\\
Imperial College, London}

\section{Introduction}

In the 15 years since Alan Guth's famous paper\cite{guth} the idea of cosmic
inflation has established itself as a central feature of modern
cosmology.  At this point the inflationary cosmologies are the only
theories which can claim to {\em explain} the initial conditions of the
Standard Big Bang Model, although people still debate the extent to
which this claim is legitimate.  One of the main difficulties
currently facing the field is the absence of a compelling microscopic
theory on which to base an inflationary cosmology.  In order to
achieve the desired period of cosmic inflation, the underlying
fundamental field theory must contain a special ``flat direction''
(the so called inflaton) with very special properties.  
A great many such field theories have been proposed, but our
understanding of very high energy physics remains sufficiently
ambiguous that there are no really strong reasons to expect the
microphyics to accommodate inflation other than our desire to explain
the ``initial conditions'' of the Universe.  

Perhaps a more serious problem for inflation is that although a given
inflationary model makes a host of very specific and calculable
predictions, a great many different models have been proposed embodying
a wide range of possible predictions.  As a result there is a
widespread perception that inflation can ``predict anything you want''
and therefor is impossible to falsify in general, even 
though specific models of inflation can be falsified.  

My goal in this contribution is to mount a defence of inflation on
this score.  Specifically, I will concentrate on the aspects of inflation
which produce ``primordial inhomogeneities'' which lead to CMB
anisotropies.  The main focus of my argument will be the predictions
which emerge from the Gaussianity and ``passivity'' of the inflationary
perturbations.  By way of contrast, I briefly mention the flatness of
the Universe and near scale invariance of the perturbation spectra as
other feature which are often discusses as ``predictions'' of inflation.

The organization of this article is as follows: Section \ref{aja-omega}
briefly comments on the question of whether 
flatness and near scale invariance are predictions of
inflation. Section \ref{aja-Gaussianity} discusses Gaussianity, and
Section \ref{aja-Passivity} presents the special predictions associated
with the ``passivity'' of inflationary perturbations.  Conclusions are 
presented in Section \ref{aja-Conclusions}.  

The basic point of this article is that when it comes to the
primordial perturbations, inflation actually commits itself to serious
constraints on the nature of these perturbations. Thus there are many
realistic experiments which can falsify the proposition that these
perturbations are predominantly inflationary {\em despite} the
enormous flexibility in the inflationary picture.  While these do not
offer the opportunity to falsify inflation altogether, such
observations would still have an enormous impact.

\section{Predictions of $\Omega$ and scale invariance}
\label{aja-omega}

It is often hotly debated whether inflation makes any concrete
predictions for $\Omega$  and for the near scale invariance of the power
spectra (for both scalar and tensor perturbations).  On one hand
inflationary models have been built which give just about any power
spectrum and $\Omega$ \cite{bt,ml}.  On the other hand certain cosmologists
have steadfastly held out for $\Omega = 1$ and a nearly scale
invariant power spectrum as strong predictions of inflation.  

I believe the intransigence on the part of those who hold out for
$\Omega = 1$ and near scale invariance as predictions, despite obvious
evidence to the contrary has conveyed an air of desperation to the
wider community. This has only enhanced the general dissatisfaction
with the predictive power of inflation.
The resulting debate has actually made inflation appear to be in a much
weaker position than is actually the case.  For one, some aspects of
inflation are falsifyable on the basis of much more concrete
predictions (as I will discuss below).  Furthermore, creating a rather
artificial debate about whether $\Omega \neq 1$ or non-scale
invariance would falsify inflation actually draws attention away from
the very substantial impact that either of these results would have.

A good analogy can be made with grand unification.  In the early days
of grand unification there was basically only one game in town:  The standard
SU(5) grand unified theory. (In fact it was this model on which the
early inflationary models were built.) When proton decay experiments
ruled out standard SU(5),  grand unification was not falsified, but
none the less the impact on the field was enormous.  It is fair to say
that the field of grand unification, with its multitude of
under-motivated models has never has regained the focus it had in the
days of standard SU(5).   

Observations that show $\Omega \neq 1$ or a not  non-scale invariant
perturbations spectrum would not rule out inflation, but they would
have an impact no less great than the proton decay experiments had on
grand unification.  The whole picture of how inflation can come about
would be severely and (by current perspective) very awkwardly
constrained.  Observers should not let philosophical discussion about
what it actually takes to falsify inflation distract them from the
importance of these observations.

\section{Gaussianity}
\label{aja-Gaussianity}

Inflation generates primordial perturbations by amplifying zero-point
quantum fluctuations to macroscopic classical scales.  Furthermore,
inflationary models which have the realistic perturbation amplitudes
have an extremely weakly coupled inflaton.  This results in a
predicted distribution of perturbations which is Gaussian
 to an excellent approximation.  

This is in fact an extremely strong prediction, and one which is easily
falsified.   A host of alternative models predict non-gaussian
features in both the matter field and the CMB anisotropies.  (Eg the
cosmic string model depicted in Figure 1.)   
\begin{figure}[t]
\centerline{\psfig{file=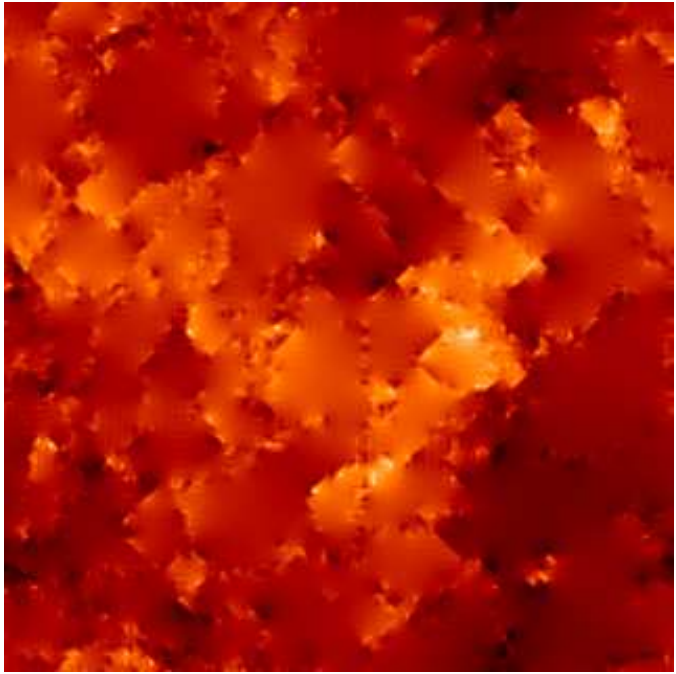,width=3.5in}}
\caption{A temperature map produced by calculating the
Kaiser-Stebbins effect for a network of cosmic strings (from [5]
).  The non-gaussian nature is clear in this approximation.}
\label{f1}
\end{figure}
A clear measurement of
any one of these characteristic signals would be strong evidence
against an inflationary origin for the perturbations.  This would
reflect a failure of the inflationary 
model which could NOT be circumvented by  fiddling around with the
inflaton potential.  
Although for certain exotic cases non-gaussianity may occur\cite{sbb}, these 
should give a very different signal which 
could not be confused with the sort of signals predicted by the defect
models. 

Given the importance of the question of Gaussianity, this is a rather
poorly developed area of research.  There are two main difficulties
which must be addressed:  

One is ``creeping Gaussianity''. The Central Limit Theorem teaches us
that the accumulated result of a lot of non-gaussian process can be
very Gaussian.  Also, a lot of practical factors in realistic
experiments (eg the presence of noise and the need for foreground
subtraction) can greatly reduce the sensitivity of an experiment to a
primordial non-Gaussian signal.  A new paper by Ferreira and
Magueijo\cite{fm} represents welcome progress in addressing these
issues, and suggests that despite the obstacles one can expect to make
realistic tests of non-Gaussianity in the CMB anisotropies. 

The second difficulty is that the term ``non-Gaussianity '' taken alone
does not mean very much.  There is a huge space of non-Gaussian
distributions, and 
a randomly chosen ``non-Gaussian statistic'' is unlikely to be
sensitive to a realistic signal.  (This has been shown in \cite{fm},
where a number of standard ``Non-Gaussian''  statistics we unable to
detect a pretty simple cosmic 
string signal.)  So the task is to determine what are ``good''
measures of non-Gaussianity which have a chance at leading to
interesting signals. While there has been reasonable progress with
(for example) the statistics of extreme peaks (``hot
spots'')\cite{ngtsim}  I feel that this is an extremely worthwhile field
of investigation which deserves more attention.  The fact that models
in which the perturbations originate from inflation can be falsified
by this kind of investigation adds motivation to this effort, and
represents a great strength of the inflationary picture.

\section{Passivity}
\label{aja-Passivity}

\subsection{Overview}

Another very distinctive feature of inflationary perturbations is the
``passivity'' of these perturbations.  ``Passivity'' refers to the
fact that perturbations are set up during the inflationary epoch and
evolve in a linear (or ``passive'') manner until gravitional collapse
goes non-linear much later in the history of the Universe.  This is in
strong contrast to ``active'' models such as defects or explosions,
which involve non-linear processes in a much more fundamental way. 
Another type of passive model is a simple standard Big Bang model
where the perturbations are put it ``by hand'' in the initial
conditions.  The evolution of perturbations in cosmologies with the
standard matter content is highly linear until late times.

The passivity of the inflationary perturbations is another feature
that cannot be ``adjusted away'' despite the great flexibility in
inflationary model building.  Furthermore, passive models give rise to
very striking signals in the angular power spectrum of CMB anisotropies
(namely the ``secondary Doppler peaks'' or ``Sakharov oscillations'').
Over the next ten years the CMB anisotropies will be mapped in
ever-increasing  detail, and the presence or absence of Sakharov
oscillations will be clearly revealed.  If the perturbations are
determined not to be passive, {\em all} inflationary models for the
origin of these perturbations will be falsified.

The differences between active and passive are quite fundamental, and
I illustrate some key aspects of these difference in the rest of this
discussion. 

\subsection{The evolution of perturbations and the CMB}

Most models of structure formation consider perturbations which
originate at an extremely early time (eg the GUT era or even the Planck
era) and which have very small amplitudes (of order $10^{-6}$) until
well into the matter era.
Perturbations of inflationary origin start as short wavelength
quantum fluctuations which evolve (during the inflationary period)
into classical perturbations on scales of astronomical interest.
Defect based models undergo a phase transition (typically at around
GUT temperatures, eg $T\approx 10^{16}GeV$) forming defects 
which generate inhomogeneities on all scales.  

For all these models, once the inflationary period and/or phase
transition is over, the Universe enters an epoch where all the matter
components obey linear equations except for the defects (if they are
present).  This ``Standard Big Bang'' epoch can be divided into three
distinct periods. The first of these is the ``tight coupling''
period where radiation and baryonic matter are tightly coupled and
behave as a single perfect fluid.  When the optical depth grows
sufficiently the coupling becomes imperfect and the 
``damping period'' is entered.  Finally there is the ``free
streaming'' period, where the CMB photons only interact with the
other matter via gravity.  While the second and third periods can have
a significant impact on the overall shape of the angular power
spectrum, all the physics which produces the Sakharov
oscillations takes place in the tight coupling regime, which is the
focus of the rest of this section.

Working in in synchronous gauge, and following the conventions and
definitions used in 
\cite{acfm}, the Fourier space perturbation
equations are:
\begin{eqnarray}
\label{eq:1}
\dot{\tau}_{00} & = & 
\Theta_{D} + 
{1 \over 2\pi G}\left( {\dot a \over a}\right)^2
\Omega_r\dot s \left[1 + R\right] \\
\dot{\delta}_c & = & 
4 \pi G {a \over \dot a}\left(\tau_{00} - \Theta_{00}\right) 
 -{\dot a \over a}\left({3\over 2} \Omega_c + 2\left[ 1+ R \right] 
\Omega_r \right)\delta_c \nonumber
\\ & & \mbox{}- {\dot a \over a}2\left[1 + R\right]\Omega_r s\\
\label{eq:3}\ddot s &= & - { \dot R \over 1 + R}\dot s - c_s^2k^2\left(s + 
\delta_c\right) \label{eq:pert}
\end{eqnarray}
Here $\tau_{\mu\nu}$ is the pseudo-stress tensor,
 $\Theta_D \equiv
\partial_i\Theta_{0i}$, 
$\Theta_{\mu\nu}$
is the defect stress energy, $a$ is the cosmic scale factor, 
$G$ is Newton's constant, $\delta_X$ is the density contrast and
$\Omega_X$ is the mean energy density over  
critical density of
species $X$ ($X=r$ for relativistic matter, 
$c$ for cold matter, $B$ for baryonic matter), $s \equiv {3\over
4}\delta_r - \delta_c$, 
$R = \frac{3}{4}\rho_B/\rho_r$, $\rho_B$ and $\rho_r$ are
the mean densities in baryonic and relativistic matter respectively, 
$c_s$ is the speed of sound and $k$ is the 
comoving wavenumber.  The dot denotes the conformal time 
derivative $\partial_\eta$.

In the inflationary case there are no defects and $\Theta_{\mu\nu}=0$.
With suitable initial conditions these linear equations completely
describe the evolution of the perturbations.  In the defect case
$\Theta_{\mu\nu}\neq 0$, and certain components
 of
$\Theta_{\mu\nu}(\eta)$ are required as input. (Conservation of stress
energy allows the equations to be manipulated so 
that different components of $\Theta_{\mu\nu}$ are required as input,
a matter mainly of numerical
convenience \cite{vs,pst,durrer,ct,ngt1-2,hw,hsw}.)
 Cosmic defects are
``stiff'', which means $\Theta_{\mu\nu}(\eta)$ can be viewed as an
external source for these equations.  The additional equations from
which one determines $\Theta_{\mu\nu}(\eta)$ are highly non-linear,
although the solutions tend to have certain scaling properties which
allow $\Theta_{\mu\nu}(\eta)$ to be modelled using a variety of
techniques \cite{ct,durrer,acfm,metal}.  

\subsection{The passive case: Squeezing and phase coherence}

Quite generically, for wavelengths larger than the Hubble radius ($R_H
\equiv a/\dot a$), Eqns [1-3] have decaying and 
growing solutions.  The growing solutions reflects the
gravitational instability, and, as required of any system which
conserves phase space volume, there is a corresponding decaying
solution.  For example, in the radiation dominated epoch, two
long wavelength solutions for the radiation perturbation
$\delta_r$ are $\delta_r \propto \eta ^{2}$ and  $\delta_r \propto
\eta ^{-2}$ (the ``adiabatic'' modes).  The adiabatic growing
solution, if present, eventually comes to dominate over any other
component.  Over time, $R_H$ grows 
compared with a comoving 
wavelength so in the Standard Big Bang epoch a given mode starts with
wavelength $\lambda >> R_H$ but eventually crosses
into the $\lambda < R_H$ regime.  In the period of tight coupling
modes with
$\lambda < R_H$ undergo oscillatory behavior since the radiation pressure stabilizes
the fluid against gravitational collapse.  This process of first
undergoing unstable behavior which eventually converts to oscillatory
behavior is the key to the formation of Sakharov oscillations.

This effect is illustrated in Figure 2, where
different solutions for $\delta_r$ are shown.  
\begin{figure}[t]
\centerline{\psfig{file=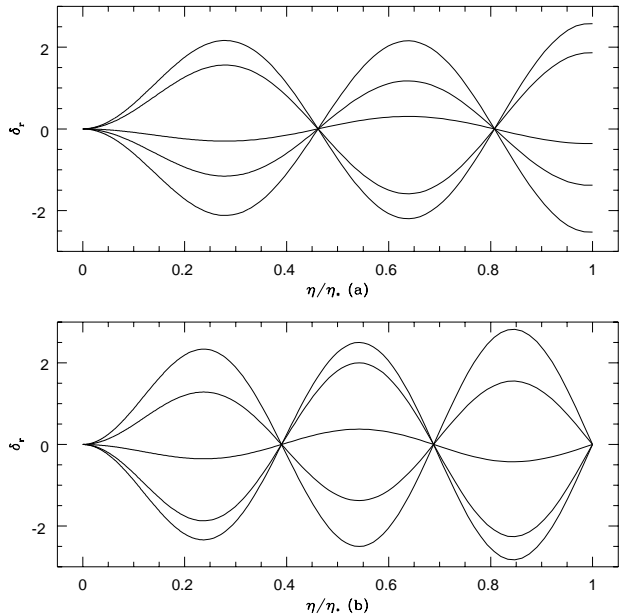,width=3.5in}}
\caption{Passive perturbations: Evolution of two different
modes during the tight 
coupling era. While
in (a) elements of the ensemble have non-zero values at $\eta_\star$,
in (b), {\it all} members of the ensemble will go to zero at
the final time ($\eta_\star$), due to the fixed phase of
oscillation set by the domination of the growing solution (or squeezing) 
which occurs before the onset of the oscillatory phase.  The
y-axis is in arbitrary units. (From [15])
}
\label{f2}
\end{figure}
The key point is that the domination of the growing solution at early
times guarantees that {\em each} member of the ensemble will match
onto an oscillating solution at late times with the {\em same}
temporal phase (for an intuitive discussion of this effect based on
simple harmonic oscillators see \cite{moriond}). 

Figure 3 shows the ensemble averaged values of $\delta$
at a fixed time as a function of wavenumber (the power spectrum).
\begin{figure}[t]
\centerline{\psfig{file=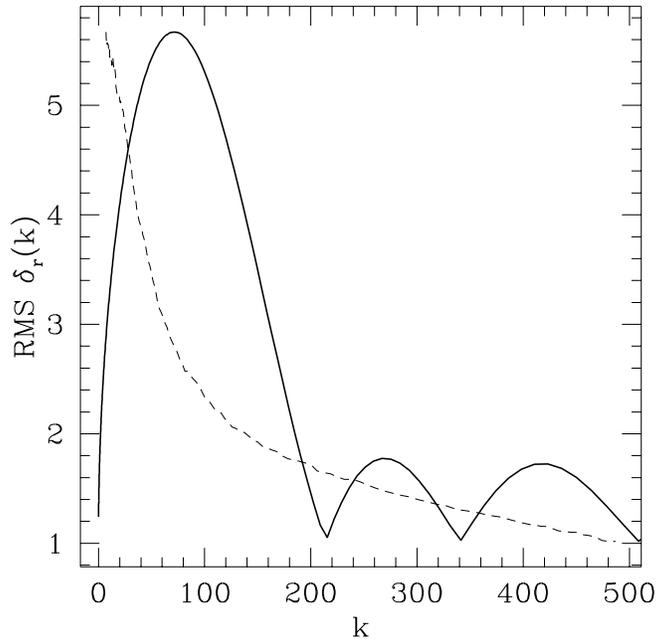,width=3.5in}}
\caption{The r.m.s.\ value of $\delta_r$ evaluated at decoupling
($\eta_\star$) for a passive model (solid) and an active model
(dashed). This is Fig 2 from ref [15]
where the details are presented.
}
\label{f3}
\end{figure}
The
zeros correspond to modes which have been caught at the ``zero-point'' of
their oscillations.  The phase focusing across the entire ensemble
guarantees that there will always be some wavenumbers where the power
is zero.  It is these zeros which are at the root of the oscillations
in the angular power spectrum.

An important point is that these zeros are an absolutely fundamental
feature in any passive theory.  No amount of tinkering with the
details can counteract that fact that an extended period of linear
evolution will lead to growing mode domination, which in turn fixes
the phase of the oscillations in the tight coupling era.   If one were
to require oscillation in a passive model to be out of phase from the
prescribed value, one would imply domination by the decaying mode
outside the Hubble radius -- in other words a Universe which is not at
all Robertson-Walker on scales greater than $R_H$.  

\subsection{The active case: Coherence lost}

As already discussed, the active case is very different from the
passive case, due to the presence of what is effectively a source term
in Eqns  [1-3].  One consequence is that the whole notion of the
ensemble average is changed.  In the passive case any model with 
Gaussian initial conditions can be solved by solving Eqns
[1-3] with the initial values for all quantities given by their
initial RMS values.  The properties of linearly  evolved
Gaussian distributions guarantee that this solution will always give
the RMS values at any time.  Thus the entire ensemble is represented
by one solution.

In the active case this is not in general possible.  One has to
average over an ensemble of possible source histories, which is a much
more involved calculation.  In \cite{metal} we ``square'' the
evolution equations to write the power spectrum as convolution of
two-point functions of the sources, but there the added
complexity requires the use of the full {\em un}equal time correlation
functions.

In general, the source term will ``drive'' the other matter
components, and temporal phase coherence will be only as strong as it
is within the ensemble of source terms.  An illustration of this
appears in Figure 4.
\begin{figure}[t]
\centerline{\psfig{file=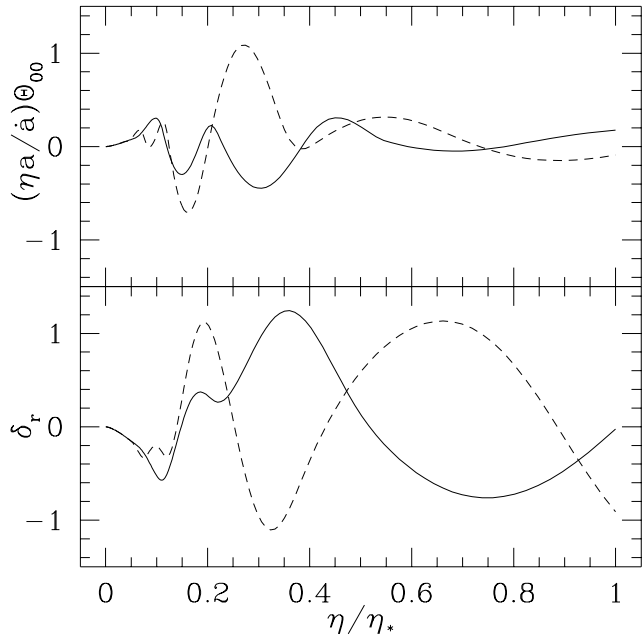,width=3.5in}}
\caption{Active perturbations:
Evolution of $\delta_r(k)$ and the corresponding source
${\Theta}_{00}$ during 
the tight  
coupling era
($\Theta_D$ is not shown).
Two members of the ensemble are shown, with matching
line types.  Due to the randomness of the source, the ensemble
includes solutions with a wide range of values at $\eta_\star$.  Unlike the
inflationary case (Figure 2) the phase of
the temporal oscillations is not fixed.  (From [15].)
}
\label{f4}
\end{figure}
In many active models the sources are sufficiently decoherent that no
oscillations appear in the power spectrum (see for example the dashed
curve in Figure 3.)

\subsection{Coherence regained}

The source evolution is a highly non-linear process, so from the
point view of a single wavenumber the source may be viewed as a
``random'' force term.  At first glance it may seem impossible for
such random force term could induce {\em any} temporal coherence, but
here is how temporal coherence can occur:  The source term
only plays a significant role in Eqns [1-3] for a {\em finite}
period of time. This is somewhat apparent in Figure 4. (The
y-axis of Figure 4 shows a quantity specially chosen to
indicate the significance of the source in Eqns [1-3].)  In
the limit where this period of significance is short compared to the
natural oscillation time of $\delta_r$ (and happens at the same time
across the entire ensemble) the ensemble of source histories {\em can}
be phase coherent.  Thus one way an active model can produce 
oscillations in the power spectrum is by having a source term which is
``sharply peaked'' in time.

It is also the case that on scales larger than $R_H$ there are
``squeezing'' mechanisms at work, even for active perturbations.  (The
gravitational instability is, after all, still present.)  In \cite{ar}
a Green function method is developed which clearly illustrates how the
active case involves a competition between squeezing effects, which
tend to produce oscillations in the power spectrum, and the
randomizing effect of the nonlinear source evolution.
In extreme cases, where the ``randomizing'' effects are minimized it
is even possible to have active models with mimic an inflationary
signal\cite{ngt1-2,hw,hsw}.

\section{Conclusions}
\label{aja-Conclusions}

I have outlined the fundamental ways in which inflationary models
commit themselves to a constrained range of possible predictions for
the primordial fluctuations. The passivity and Gaussianity of
inflationary models leads to strong predictions which can be
falsified by experiments which 
will be completed within a decade.  Thus, despite the numerous
uncertainties in inflationary model building, the inflationary
scenarios present an opportunity to do absolutely first rate science. 

What if an inflationary origin for the perturbations is ruled out?
The idea of inflation as a solution for other cosmological problems
will still be attractive.  In fact, it has been argued that perhaps
the most ideal role for inflation is to solve the horizon and flatness
problems while producing negligible perturbations\cite{toa,vilenkin},
leaving the origin of perturbations to some 
alternative like cosmic defects, which have their own unique signals.
Either way, the new data is certain to have a profound impact on the
world of theoretical cosmology.

\section*{Bibliographic Notes} 


\begin{thebibliography}{}   

\bibitem{guth} A. Guth, {\it Phys. Rev. D} {\bf 23} 347, (1981)
\bibitem{bt} M. Bucher and N. Turok, hep-ph/9503393, (1995) M. Bucher,
A. S. Goldhaber, N. Turok, hep-ph/9501396, (1995).
\bibitem{ml} A. Linde, and A. Mezhlumian, {\em Phys. Rev. D} {\bf 52},
6789 (1995).
\bibitem{sbb} D.S. Salopek, J.R. Bond and J. M. Bardeen, {\em
Phys. Rev. D} {\bf 40}, 1753 (1989), A. Linde and V. Mukhanov,
astro-ph/9610219 (1996) M. Bucher and Y. Zhu, astro-ph/9610223 (1996).
\bibitem{fm} P. Ferreira and J. Magueijo, astro-ph/961017, 1996
\bibitem{ngtsim} N. Turok, astro-ph/9606087 (1996)
\bibitem{acfm} 
A. Albrecht, D. Coulson, P. Ferreira, and J. Magueijo,
{\it Phys. Rev. Lett.} {\bf 76} 1413-1416 (1996).
\bibitem{vs}
S. Veeraraghavan and A. Stebbins, {\it Astrophys.J.}
{\bf 365} 37-65 (1990).
\bibitem{pst}U. Pen, D.N. Spergel and N. Turok, {\it Phys. Rev. D},
{\bf 49} 692-729 (1994).
\bibitem{durrer}R. Durrer, A. Gangui and M. Sakellariadou, 
{\it Phys. Rev. Lett.}, {\bf 76} 579 (1996).
\bibitem{ct}
R. Crittenden and N. Turok, {\it Phys. Rev. Lett.}, {\bf 75}
2642 (1995).
\bibitem{ngt1-2} N. Turok, DAMTP-preprint-96-44, astro-ph/9604173 (1996),
 N. Turok, DAMTP-preprint-96-69, astro-ph/9607109
(1996).
\bibitem{hw} W. Hu and M. White, IAS preprint astro-ph/9603019 (1996)
\bibitem{hsw} W. Hu. D. Spergel, and M. White IAS preprint
astro-ph/9605193 (1996) 
\bibitem{metal} 
J.Magueijo, A. Albrecht, D. Coulson, P. Ferreira, {\it Phys. Rev. Lett},
{\bf 76} 2617 (1996), 
J.Magueijo, A. Albrecht, P. Ferreira, D. Coulson, {\it Phys. Rev. D},
{\bf 54}, 3727, (1996)

\bibitem{moriond} A. Albrecht. In {\it The proceeding of the Moriond
Meeting on CMB Anisotropies} March 1996 (Imperial/TP/96-97/8).

\bibitem{ar} A. Albrecht and J. Robinson, in preparation (1996).

\bibitem{toa} A. Albrecht, in the proceedings of {\em The international workshop on the Birth of the Universe and
Fundamental Forces} Rome, 1994 F. Occionero ed. (Springer Verlag) 1995.

\bibitem{vilenkin}
A.~Vilenkin. {\it Phys. Rev. Lett.} {\bf 74}  846-849, (1995)

\end{thebibliography}
\end{document}